\documentclass[10pt]{article}
\usepackage{amsmath}
\usepackage{amssymb}
\usepackage{amsthm}
\pdfoutput=1

\usepackage[utf8]{inputenc}
\usepackage[T1]{fontenc}
\usepackage{microtype}

\usepackage{fourier}
\usepackage{caption}
\usepackage{mathrsfs}
\usepackage[colorlinks=true, pdfstartview=FitV, linkcolor=blue, citecolor=blue, urlcolor=blue]{hyperref}
\theoremstyle{plain}
\newtheorem{proposition}{Proposition}[section]

\theoremstyle{definition}

\newtheorem{example}[proposition]{Example}

\usepackage[usenames,dvipsnames]{xcolor}
\usepackage{url}
\usepackage{tikz}
\usepackage{pgfplots}
\pgfplotsset{compat=newest,ticks=none}
\usetikzlibrary{arrows,calc}
\usepackage{verbatim}
\usetikzlibrary{%
    decorations.pathreplacing,%
    decorations.pathmorphing%
}

\newcommand{\R}{\mathbf{R}}
\newcommand{\C}{\mathbf{C}}

\newcommand{\abs}[1]{\left\lvert #1 \right\rvert}

\usepackage{mathscinet}
\usepackage{cite}

\usepackage{microtype}
\begin{document}

\title{An inverse problem for the modified Camassa-Holm equation and multi-point Pad\'{e} approximants}

\author{Xiang-ke Chang\thanks{Department of Mathematics and Statistics, University of Saskatchewan, 106 Wiggins Road, Saskatoon, Saskatchewan, S7N 5E6, Canada; changxk@lsec.cc.ac.cn}
\and
  Jacek Szmigielski\thanks{Department of Mathematics and Statistics, University of Saskatchewan, 106 Wiggins Road, Saskatoon, Saskatchewan, S7N 5E6, Canada; szmigiel@math.usask.ca}
}

%\date{June 16, 2012}
\date{\today}

\maketitle

\begin{abstract}
  In this Letter the main steps in the inverse spectral construction of a family of non-smooth solitons ~(\emph{peakons})
  to the modified Camassa-Holm equation are oulined.
  It is shown that the inverse problem is solvable in terms of multi-point
  Pad\'{e} approximations.

\end{abstract}


% 34  ODE
% 34B Boundary value problems
% 34K Functional-differential and differential-difference equations

% 35  PDE
% 35Q Equations of mathematical physics and other areas of application

% 37  Dynamical systems and ergodic theory
% 37J Finite-dimensional Hamiltonian, Lagrangian, contact, and nonholonomic systems

% 41  Approximations and expansions

%\subjclass[2010]{Primary 34A55} % 34A55 Ordinary differential equations - General theory - Inverse problems

%\tableofcontents

%\section{Introduction}

%\section{}

\vspace{0.5 cm}

The nonlinear partial differential equation
\begin{equation}\label{eq:m1CH}
m_t+\left((u^2-u_x^2) m)\right)_x=0, \qquad
m=u-u_{xx},
\end{equation}
is an intriguing modification of the Camassa-Holm equation
~$m_t+u m_x +2u_x m=0, \, ~m=u-u_{xx}$ \cite{CH}, for the shallow water waves.
Originally, equation \eqref{eq:m1CH} appeared in the papers of Fokas \cite{fokas1995korteweg}, Fuchssteiner \cite{fuchssteiner1996some},  Olver and Rosenau\cite{olver1996tri}. As mentioned by Fokas, this equation can also arise in the theory of nonlinear water waves.
%and was, later, rediscovered by Qiao \cite{qiao2006new,qiao2007new} who derived this equation as a formal approximation to the two dimensional Euler equation for an inviscid, incompressible, irrotational fluid; the fact also mentioned in passing in \cite{fokas1995korteweg}.
We are interested in the class of non-smooth solutions given by the peakon ansatz \cite{CH,qiao2012integrable,gui2013wave}, that is, we assume $u=\sum_{j=1}^n m_j (t)e^{-\abs{x-x_j(t)}}, $ ~$x_1<x_2< \dots < x_n$, all $m_j$s are positive and hence $ m=2\sum_{j=1}^n m_j \delta_{x_j}$ is a positive discrete measure.  This forces us to view \eqref{eq:m1CH} as a distributional equation which requires that we
define the product $u_x^2 m$.  We will argue elsewhere that the
choice consistent with integrability is to take $u_x^2 m$ to mean $\langle u_x^2 \rangle m$,
where $\langle f \rangle$ denotes the average function (the arithmetic average
of the right hand and left hand limits). Subsequently, equation \eqref{eq:m1CH}
reduces to the system of ODEs:
\begin{equation}\label{eq:xmODE}
\dot m_j=0, \qquad \dot x_j=u(x_j)^2-\langle u_x(x_j)^2 \rangle.
\end{equation}
We note that this system is not exactly the same as the one proposed in \cite{gui2013wave,qiao2012integrable};
the difference being precisely in the definition of the singular product $u_x^2 m$.
In broad terms we can say that our general interest in \eqref{eq:m1CH} is to understand how integrability manifests itself in the non-smooth sector of solutions, in particular how it determines the properties of such singular operations.

The Lax
pair for \eqref{eq:m1CH} reads \cite{qiao2006new}:
$\Psi_x=\frac12U \Psi, \quad  \Psi _t =\frac12 V \Psi, \quad  \Psi=\begin{bmatrix} \Psi_1\\\Psi_2 \end{bmatrix} $ with
\begin{equation*}
U=\begin{bmatrix} -1 &\lambda m\\ -\lambda m& 1 \end{bmatrix}, \qquad
V=\begin{bmatrix} 4\lambda^{-2} + Q & -2\lambda^{-1} (u-u_x)-\lambda m Q\\
2\lambda^{-2}(u+u_x)+\lambda m Q & -Q \end{bmatrix}, \quad Q=u^2-u_x^2.
\end{equation*}
Performing the gauge transformation $\Phi=\textrm{diag}(e^{\frac x2}, e^{-\frac x2}) \Psi$ results in a simpler $x$-equation
\begin{equation}\label{eq:xLax}
\Phi_x=\begin{bmatrix}0 &\lambda h\\
-\lambda g& 0 \end{bmatrix} \Phi, \qquad    g=\sum_{j=1} ^ng_j \delta_{x_j}, \qquad h=\sum_{j=1} ^nh_j \delta_{x_j},
\end{equation}
where $g_j=m_j e^{-x_j}, \, h_j =m_j e^{x_j}$.
We will be interested in solving the boundary value problem \eqref{eq:xLax}
with boundary conditions $\Phi_1(-\infty)=0, \, \Phi_2(+\infty)=0$.  To make the
boundary value problem well defined we require that $\Phi$ be left continuous and we define the terms $\Phi_{a}\delta_{x_j}=\Phi_a(x_j)\delta_{x_j}, a=1,2$.  This choice makes the Lax pair well defined as a distributional Lax pair and
the compatibility condition indeed implies \eqref{eq:xmODE}.  We furthermore remark that there is no nontrivial solution to this boundary value problem if $\lambda=0$.
The solution $\Phi$ is a piecewise constant function which, for convenience, we can normalize by setting $\Phi_2(-\infty)=1$.  The distributional boundary value problem is in our special case of the discrete measure $m$ equivalent to a finite difference equation. Indeed, if we define $q_k=\frac{\Phi_1(x_{k})}{\lambda}, \, ~p_k=~\Phi_2(x_{k}), \, z=~\lambda^2$ then the difference form of the
boundary value problem reads:
\begin{equation} \label{dstring}
\begin{gathered}
\begin{aligned}
 q_{k+1}-q_{k}&=h_kp_k, & 1\leq k\leq n, \\
 p_{k+1}-p_k&=-z g_kq_{k},& 1\leq k\leq n,\\
  q_1=0, \quad  p_1=1&, \quad p_{n+1}=0, &
  \end{aligned}
\end{gathered}
\end{equation}
where $p_{n+1}=\lim_{x\rightarrow+\infty}\Phi_2(x), \, q_{n+1}=\lim_{x\rightarrow+\infty}{\frac{\Phi_1(x)}{\lambda}}$.
Solving the recurrence relations \eqref{dstring} by induction shows that $q_{k+1}(z)$ is a polynomial of degree $\lfloor\frac{k-1}{2}\rfloor$ in $z$, and $p_{k+1}(z)$ is a polynomial of degree $\lfloor\frac{k}{2}\rfloor$, respectively. It follows that the cardinality of the spectrum of the boundary value problem (\ref{dstring})  is at most $\lfloor \frac n2 \rfloor$; in fact it is exactly $\lfloor \frac n2 \rfloor$ because, as we will argue below, the spectrum is simple (and positive).

For $x>x_{n}$,  the $t$ part of the Lax pair implies
\begin{equation}\label{eq:tderqp}
  \dot q_{n+1}=\frac{2}{z}q_{n+1}-\frac{2L}{z}\,p_{n+1}, \qquad \dot p_{n+1}=0,
\end{equation}
where $L=\sum_{j=1}^{n}h_j$. Thus $p_{n+1}(z)$ is independent of time and its zeros, i.e. the spectrum, are time invariant.

Note that the difference form of the boundary value problem admits two natural  matrix presentations the first one of which is just the $2\times 2$ matrix encoding of \eqref{dstring}
\begin{equation}\label{transition}
\begin{bmatrix}
  q_{k+1}\\
  p_{k+1}
\end{bmatrix}
=T_k\begin{bmatrix}
  q_k\\
  p_{k}
\end{bmatrix}, \qquad\qquad
T_k=\begin{bmatrix}
  1& h_k\\
  -z g_k&1
\end{bmatrix}.
\end{equation}
We point out that the transition matrix $T_k$ is significantly different from the difference equation for the inhomogeneous
string boundary value problem discussed in \cite{lundmark2005degasperis} (Appendix A) which has
$T_k=\begin{bmatrix}
  1& l_{k-1}\\
  -z g_k&1-zg_kl_{k-1}
\end{bmatrix}, $
thus an element of the group $SL_2(\C)$.
The second formulation of the boundary value problem is in terms of $(n-1)\times (n-1)$ matrices.  Indeed, let
\begin{equation*}
\mathbf{q}=\begin{bmatrix}q_2\\q_3\\\vdots\\q_n \end{bmatrix}, \qquad
 \mathbf{p}=\begin{bmatrix}p_2\\p_3\\ \vdots\\p_n \end{bmatrix}, \qquad
 G= \textrm{ diag } (g_2,g_3,\dots, g_n), \qquad H=\begin{bmatrix}h_1&0&\dots &\dots&\dots&0\\
 h_2&0&0&\dots&\dots&0\\
 0&h_3&0&\dots&\dots&0\\
 0&0&h_4&0&\dots &0\\
 \vdots&\vdots&\vdots&\vdots&0\\
 0&\hdots&0&&h_{n-1}&0 \end{bmatrix},
 \end{equation*}
 and let $E=[\delta_{i, j-1}], E^T$ denote the unilateral shift matrix and its transpose
 respectively.
 Then \eqref{dstring} implies
 \begin{equation*}
\mathbf p=z(I-E)^{-1}G(I-E^T)^{-1} H \mathbf p\stackrel{\textrm{def}}{=}zK \mathbf p.
\end{equation*}
Since in our case $h_j>0, g_j>0$, $K$ is an irreducible totally non-negative matrix and all its non-zero eigenvalues
are positive and simple by a result in \cite{Gekhtman}.  Hence all non-zero
reciprocals of eigenvalues, say $\zeta_j$, solve $\det (I-zK)=0$ and, after checking the normalization, we conclude that $p_{n+1}(z)=\det(I-zK)$, which in turn implies that the degree of $\det(I-zK)$
is exactly $\lfloor \frac n2 \rfloor$.  We conclude this brief survey of the
forward spectral map by introducing the Weyl function for the boundary value problem
\eqref{dstring}:
\begin{equation*}
W(z)=\frac{q_{n+1}(z)}{p_{n+1}(z)},
\end{equation*}
which, in view of equations \eqref{eq:tderqp}, undergoes a linear evolution
$\dot W=\frac{2}{z}W-\frac {2L}{z}$.  Since the spectrum is simple and $\textrm{deg}q_{n+1}\leq \textrm{deg}p_{n+1}$,
$W(z)=\sum_{j=1}^{\lfloor \frac n2 \rfloor} \frac{b_j}{\zeta_j-z}+c$ where
$b_j(t)=b_j(0)e^{\frac{2\, t}{\zeta_j}}, c=c(0)$.  Moreover, as it is demonstrated below, all residues satisfy $b_j(0)>0$ and $c\geq0$; more precisely $c>0$ when $n$ is odd
and $c=0$ for $n$ even.  Thus $W(z)$ can be written
\begin{equation}\label{eq:WrepS}
W(z)\stackrel{\textrm{def} }{=}c_{2n}+\int \frac{d\mu_{2n}(x)}{x-z}, \qquad d\mu_{2n}=\sum_{j=1}^{\lfloor \frac n2 \rfloor} b_j^{(2n)}\delta_{\zeta_j},
\end{equation}
where the extra index $2n$ is introduced to facilitate the formulation of the inverse problem.

The inverse problem can be stated: given positive constants $m_j,\, 1\leq j\leq n$, and a rational function $W(z)$ with integral representation \eqref{eq:WrepS}, we seek to invert the map $S: \{x_1, x_2, \dots x_n\}\longrightarrow W$.  To solve the
inverse problem we proceed in two stages: first we
reconstruct positive coefficients $g_j, h_j$ such that $g_jh_j=m_j^2$ then
we use the relation $\frac{h_j}{g_j}=e^{2x_j}$ to determine $x_j$.
The reconstruction of $h_j,g_j$ amounts to solving a recurrence relation
following the steps below:
\begin{enumerate}
\item starting with $w_{2n}=W(z)$ define $h_n=w_{2n}(-\frac{1}{m_n^2}), \, \, g_n=\frac{m_n^2}{h_n}$ and solve
\begin{equation*}
w_{2n}=(1+zm_n^2)w_{2n-1}+h_n, \qquad \frac{1}{w_{2n-2}}=\frac{1}{w_{2n-1}}+zg_n,
\end{equation*}
for $w_{2n-1}$ and $w_{2n-2}$;
\item restart the procedure from $w_{2n-2}$ shifting $n\rightarrow n-1$.
\end{enumerate}
We remark that the procedure encodes solving \eqref{dstring}
backwards using $w_{2j}=\frac{q_{j+1}}{p_{j+1}}, w_{2j-1}=\frac{q_{j}}{p_{j+1}}$.
However, for the procedure to make sense, $w_{2n-2}$ needs to be of the form
\eqref{eq:WrepS}.  To see that this is indeed the case we use some old results
of Stieltjes, appropriately adapted to our setup.  More specifically, it follows from general results proved in \cite{stieltjes} that any rational function $F(z)$ satisfying the spectral
representation of the kind referred to in \eqref{eq:WrepS} has a continued fraction expansion
\begin{equation*}
F(z)=c+\int \frac{d\mu(x)}{x-z}=c+\cfrac{1}{a_1 (-z)+\cfrac{1}{a_2+\cfrac{1}{a_3(-z)+\cfrac{1}{\ddots}}}}
\end{equation*}
where all $a_j>0$ and conversely any rational function with this type of
continued fraction expansion has a spectral representation \eqref{eq:WrepS}.
Now we turn to analyzing $w_{2n-2}$.
First, from the recurrence relation we easily get $
h_n=c_{2n}+\int\frac{d\mu_{2n}(x)}{x+\frac{1}{m_n^2}}=c_{2n}+m_n^2\int d\mu_{2n-1}(x)$ while solving for $w_{2n-1}$ yields $w_{2n-1}(z)=\int \frac{d\mu_{2n-1}(x)}{x-z}$, where
$d\mu_{2n-1}=\frac{d\mu_{2n}}{1+m_n^2 x}$, and thus
\begin{equation*}
w_{2n-1}(z)=\cfrac{1}{a_1 (-z)+\cfrac{1}{a_2+\cfrac{1}{a_3(-z)+\cfrac{1}{\ddots}}}}
\end{equation*}
for some $a_j>0$. Next, we write $$w_{2n-2}=\cfrac{1}{zg_n+\cfrac{1}{w_{2n-1}}}=\cfrac{1}{(g_n-a_1)z+\cfrac{1}{a_2+\cfrac{1}{a_3(-z)+\cfrac{1}{\ddots}}}}$$ and observe that for $w_{2n-2}$
to have the spectral representation \eqref{eq:WrepS} $g_n-a_1$ must be negative or
$0$.  However, $\frac{1}{a_1}-\frac{1}{g_n}=\int d\mu_{2n-1}(x)-\frac{h_n}{m_n^2}=
-\frac{c_{2n}}{m_n^2}$, hence $g_n-a_1 \leq 0$, which proves the existence of
the spectral representation \eqref{eq:WrepS} for $w_{2n-2}$ for some
measure $d\mu_{2n-2}$ supported on a finite number of points in $\R_+$.

It is helpful to point out the following dichotomy: if $c_{2n}=0$, which happens whenever $n$ is even, then
the support of $w_{2n-2}$ has one less point in the spectrum of the corresponding
measure compensated by the appearance of non-zero $c_{2n-2}$.  If, on the other
hand, $n$ is odd, in which case $c_{2n}>0$, then $c_{2n-2}=0$ and the number of points in the support of $d\mu_{2n-2}$ does not differ from that of $d\mu_{2n}$.
In either case, by iterating, one reaches $w_2$ which is a positive constant equal by definition to $h_1$ and the iteration stops.  Moreover, the iteration can
be reversed, that is starting from the set of $\{h_1,h_2, \dots, h_n, m_1,m_2, \dots, m_n\}$ of strictly positive numbers, we can construct $w_2, w_3, \dots w_{2n}$
with the initial input $w_2=h_1$ that will solve the recurrence relation.
This in particular proves that the spectral measure $d\mu_{2n}$ is a positive
measure supported on $\lfloor \frac n2 \rfloor$ distinct positive numbers.

The iteration proposed above requires $2n-2$ steps to reach $w_2$,
each step leading to a new input rational function $w_j$.  All these functions
are rational functions of the initial $W=w_{2n}$ and the variable $z$.  The formulas
for $h_j$ get increasingly more complicated and a natural
question presents itself: can one compute $h_j$ using directly the spectral data
$c_{2n}$ and $d\mu_{2n}$?  The answer is affirmative: there exist
determinantal formulas for $h_j$ and the origin of these formulas can be traced back to certain multipoint Pad\'{e} approximation problem.
We leave the details to a longer publication but provide the heuristics
of the approximation problem.  To this end we rewrite \eqref{transition} in terms of the Weyl function $W=w_{2n} $ as
\begin{equation} \label{eq:Papprox1}
\begin{bmatrix} W(z) \\ 1 \end{bmatrix} =T_n(z)T_{n-1}(z)  \dots T_{n-j}(z)\begin{bmatrix} \frac{q_{n-j}(z)}{p_{n+1}(z)}\\ \frac{p_{n-j}(z)}{p_{n+1}(z)} \end{bmatrix}.
\end{equation}
Clearly, the transpose of the matrix of cofactors of each $T_j(z)$
is $\begin{bmatrix} 1 &- h_j\\zg_j &1 \end{bmatrix}\stackrel{\textrm {def}}{=}C_j(z)$,
which allows one to express equation \eqref{eq:Papprox1}  as
\begin{equation*}
C_{n-j}(z)C_{n-j+1}(z)\dots C_n(z)\begin{bmatrix} W(z) \\ 1 \end{bmatrix} =\det (T_n(z))\det (T_{n-1}(z))  \dots \det(T_{n-j}(z))\begin{bmatrix} \frac{q_{n-j}(z)}{p_{n+1}(z)}\\ \frac{p_{n-j}(z)}{p_{n+1}(z)} \end{bmatrix},
\end{equation*}
which, recalling that $\det T_j(z)=1+zm_j^2$, implies
\begin{equation} \label{eq:Paprrox2}
\left( C_{n-j}(z)C_{n-j+1}(z)\dots C_n(z)\begin{bmatrix} W(z) \\ 1 \end{bmatrix}\right)\Big |_{z=-\frac{1}{m_{n-i}^2}}=0,  \qquad \textrm{ for any } 0\leq i\leq j.
\end{equation}
If we denote the matrix of products of $C$s by $\begin{bmatrix} a_j(z)&b_j(z)\\c_j(z)&d_j(z) \end{bmatrix}$ then the polynomials $a_j(z), b_j(z),c_j(z),d_j(z)$
solve the following interpolation problem:
\begin{subequations}\label{eq:Papprox}
\begin{align}
&a_j(-\frac{1}{m_{n-i}^2})W(-\frac{1}{m_{n-i}^2})+b_j(-\frac{1}{m_{n-i}^2})
=0, \qquad 0\leq i\leq j, \\
&\deg a_j=\big\lfloor \frac{j+1}{2} \big \rfloor, \qquad \deg b_j=\big \lfloor \frac j2 \big\rfloor, \qquad a_j(0)=1, \\
\notag\\
&c_j(-\frac{1}{m_{n-i}^2})W(-\frac{1}{m_{n-i}^2})+d_j(-\frac{1}{m_{n-i}^2})
=0, \qquad 0\leq i\leq j,  \\
&\deg c_j=\big \lfloor \frac{j}{2}\big \rfloor +1, \qquad \deg d_j=\big \lfloor \frac{j+1}{2}\big \rfloor, \qquad c_j(0)=0, \quad d_j(0)=1,
\end{align}
\end{subequations}
which is an example of a \textit {Cauchy-Jacobi interpolation problem} \cite{gonvcar1978markov,meinguet1970on,stahl1987existence} studied as part of a general multi-point Pad\'{e} approximation theory \cite{baker1996pade}.

In the second stage of solving the inverse problem we need to
solve for $x_j=\frac 12 \ln \frac{h_j}{g_j}$.  Since both $h_j, g_j$ are strictly positive
the solution exists but it may violate the ordering condition $x_1<x_2<\dots <x_n$.
In the forthcoming paper we propose a set of sufficient conditions on $m_i$s which depend only on the spectrum but not on the residues $b_j$ thus ensuring the global in $t$ existence of
peakon solutions described in this paper.

We conclude this Letter by citing a few computational results from
the forthcoming  paper in which we provide the details of the complete
solution to the peakon problem for the modified CH equation \eqref{eq:m1CH}. We also highlight sufficient conditions for the global existence of solutions. We recall
that the only time dependence is in the evolution of the residues:
$b_j(t)=b_j(0)e^{\frac{2\, t}{\zeta_j}}$.
\begin{example}[2-peakon solution: $\frac{1}{\zeta_1}<m_1m_2$]
  \begin{align*}
     &x_1=\ln\left(\frac{b_1}{\zeta_1m_1(1+\zeta_1m_2^2)}\right),\ \ \
     &x_2=\ln\left(\frac{b_1m_2}{1+\zeta_1m_2^2}\right).
       \end{align*}

\end{example}

\begin{example}[4-peakon solution: $\frac{1}{\zeta_1}<m_1m_2, \, \, \frac{\zeta_2}{\zeta_1^2}<m_3m_4, \, \, \frac{m_2 m_3}{(1+\zeta_1 m_2^2)(1+\zeta_ 1m_3^2)}<
\frac{\zeta_1}{(\zeta_2-\zeta_1)^2}$]
  \begin{align*}
     &x_1=\ln\left(\frac{1}{m_1}\cdot\frac{b_1b_2(\zeta_2-\zeta_1)^2}{\zeta_1\zeta_2\left(b_1\zeta_1(1+\zeta_2m_2^2)(1+\zeta_2m_3^2)(1+\zeta_2m_4^2)+b_2\zeta_2(1+\zeta_1m_2^2)(1+\zeta_1m_3^2)(1+\zeta_1m_4^2)\right)}\right),&\\
     &x_2=\ln\left(m_2\cdot\frac{b_1b_2(\zeta_2-\zeta_1)^2\left(b_1(1+\zeta_2m_3^2)(1+\zeta_2m_4^2)+b_2(1+\zeta_1m_3^2)(1+\zeta_1m_4^2)\right)}{\left(b_1\zeta_1(1+\zeta_2m_3^2)(1+\zeta_2m_4^2)+b_2\zeta_2(1+\zeta_1m_3^2)(1+\zeta_1m_4^2)\right)}\right.&\\
     &\qquad\qquad \cdot\left.\frac{1}{\left(b_1\zeta_1(1+\zeta_2m_2^2)(1+\zeta_2m_3^2)(1+\zeta_2m_4^2)+b_2\zeta_2(1+\zeta_1m_2^2)(1+\zeta_1m_3^2)(1+\zeta_1m_4^2)\right)}\right),&\\
     &x_3=\ln\left(\frac{1}{m_3}\cdot\frac{\left(b_1(1+\zeta_2m_4^2)+b_2(1+\zeta_1m_4^2)\right)\left(b_1(1+\zeta_2m_3^2)(1+\zeta_2m_4^2)+b_2(1+\zeta_1m_3^2)(1+\zeta_1m_4^2)\right)}{(1+\zeta_1m_4^2)(1+\zeta_2m_4^2)\left(b_1\zeta_1(1+\zeta_2m_3^2)(1+\zeta_2m_4^2)+b_2\zeta_2(1+\zeta_1m_3^2)(1+\zeta_1m_4^2)\right)}\right),&\\
     &x_4=\ln\left(m_4\cdot\frac{b_1(1+\zeta_2m_4^2)+b_2(1+\zeta_1m_4^2)}{(1+\zeta_1m_4^2)(1+\zeta_2m_4^2)}\right).&
       \end{align*}
\end{example}
\bibliographystyle{abbrv}
%\bibliography{IP,SZ,peakon508,IS508,OPandIS508,hankel508}
\def\cydot{\leavevmode\raise.4ex\hbox{.}}
  \def\cydot{\leavevmode\raise.4ex\hbox{.}}

\end{document}